# Final excitation energy of fission fragments

## Karl-Heinz Schmidt and Beatriz Jurado

### CENBG, CNRS/IN2P3, Chemin du Solarium, B. P. 120, 33175 Gradignan, France

**Abstract:** We study how the excitation energy of the fully accelerated fission fragments is built up. It is stressed that only the intrinsic excitation energy available before scission can be exchanged between the fission fragments to achieve thermal equilibrium. This is in contradiction with most models used to calculate prompt neutron emission where it is assumed that the total excitation energy of the final fragments is shared between the fragments by the condition of equal temperatures. We also study the intrinsic excitation-energy partition according to a level density description with a transition from a constant-temperature regime to a Fermi-gas regime. Complete or partial excitation-energy sorting is found at energies well above the transition energy.

**PACS: 24.75.+i, 24.60.Dr, 21.10.Ma**

**Introduction:**

The final excitation energy found in the fission fragments, that is, the excitation energy of the fully accelerated fission fragments, and in particular its variation with the fragment mass, provides fundamental information on the fission process as it is influenced by the dynamical evolution of the fissioning system from saddle to scission and by the scission configuration, namely the deformation of the nascent fragments. The final fission-fragment excitation energy determines the number of prompt neutrons and gamma rays emitted. Therefore, this quantity is also of great importance for applications in nuclear technology. To properly calculate the value of the final excitation energy and its partition between the fragments one has to understand the mechanisms that lead to it. In particular, one has to take into account that the different contributions to the total excitation energy (*TXE)* of the fission fragments appear at different stages of the fission process.

The intrinsic excitation energy of the fissioning system acquired before scission, $E^*_{intr}$, may be exchanged between the nascent fragments. This process strongly influences the prompt neutron yield as a function of fragment mass. The second law of thermodynamics drives the system towards thermal equilibrium. The energy partition will be the one that maximizes the entropy of the system made of the two fragments in contact near scission. Consequently, to investigate the sharing of $E^*_{intr}$ one has to consider the level density of the emerging fission fragments before scission. Since long, there have been experimental indications that the nuclear level density at low excitation energies shows a constant-temperature behavior. This has already been realized by Gilbert and Cameron [1]. The constant-temperature behavior deviates from the Bethe formula [2], also named Fermi-gas level density, which predicts a behavior of the level density like $e^{2\sqrt{aE^*}}$ in the independent-particle model (*a* is the level-density parameter). A suggestive solution of this apparent contradiction is the influence of residual interactions (e.g. pairing correlations), which explains the deviations from the independent-particle model at lower excitation energies. This idea led already Gilbert and Cameron to propose their composed level-density formula that describes the level density by a constant-temperature formula below and the back-shifted Fermi-gas formula (the Bethe formula with an energy shift) above a "matching energy" [3]. The value of the matching energy is determined by the assumption that there is a sudden transition from one description to the other and by the condition that the two descriptions coincide at the matching energy. Rather recently, new experimental results [4] suggest that the validity of the constant-temperature regime extends to appreciably higher excitation energies than the matching energy according to the composed Gilbert-Cameron description. Thus, the transition between the constant-temperature regime and the independent-particle regime is subject to current research. Still, the theoretical arguments for the validity of the independent-particle model at higher excitation energy remain valid.

In a previous work, we showed that thermodynamic processes lead to a complete energy sorting of the thermal excitation energy of the nascent fragments before scission in the constant-temperature regime [5]. In another paper, we discussed the energy sharing in the regime of the back-shifted Fermi-gas description [6]. In the present work, we investigate the sharing of excitation energy between the two nascent fragments at energies around the matching energy in the composed Gilbert-Cameron description. The relevance of this study is not only given by the fact that the Gilbert-Cameron description is used rather frequently for the modeling of nuclear properties and nuclear reactions for technical applications, but even more by the expectations that the transition from a constant-temperature-like behavior at low excitation energies to an independent-particle description at higher excitation energies is a realistic feature of the nuclear level density. The essential of the present study is independent of the exact value of the transition energy, and the results may easily be adapted to an extended validity range of the constant-temperature regime.

**Contributions to the final fission-fragment excitation energy**

The excitation energy of the fully accelerated fragments is composed of the following contributions: (i) the intrinsic excitation energy, (ii) the excitation energy stored in collective excitations, (iii) the deformation energy (mostly due to the larger surface of the strongly deformed nascent fragments at scission) with respect to the ground state of the fragment. Figure 1 illustrates that these different contributions to the *TXE* appear at different stages of the fission process.

The intrinsic excitation energy $E^*_{intr}$ at scission is given by:

$$E^*_{intr} = E^*_{CN} - FB + E_{dis} \qquad (1)$$

where $E^*_{CN}$ is the initial energy of the fissioning nucleus. For example in neutron-induced reactions it is given by the sum of the neutron binding energy and the incident neutron energy in the centre-of-mass reference system. *FB* is the height of the fission barrier and $E_{dis}$ is the amount of the energy released on the way from saddle to scission that is dissipated into intrinsic excitation energy. Let us present an example to give a quantitative idea of the amount of $E^*_{intr}$. The measured prompt-neutron yields shown in Fig. 1 of Ref. [5] ($^{237}$Np(n,f) with $E_n$ = 0.8 and 5.5 MeV) correspond to an excitation energy of 0.1 MeV, respectively 4.8 MeV, above the fission barrier. According to Ref. [7], the energy release from saddle to scission in $^{238}$Np is about 17 MeV. Theoretical estimations by e.g. [8] state that only a minor part of this energy is dissipated into intrinsic excitation energy at scission. Thus, the intrinsic excitation energies at scission, which are relevant for the discussion in Ref. [5], are expected to be around 10 MeV, certainly well below 20 MeV. Of course, neutron energies above 5.5 MeV would lead to higher intrinsic excitation energies.

The intrinsic excitation energy $E^*_{intr}$ that is present in the fissioning system before scission is divided between the fission fragments, e.g. according to thermal equilibrium. The excitation energy stored in one of the collective normal modes (angular-momentum bearing and others) [9] is shared between the fragments according to the corresponding coordinated motions of the nascent fragments. It is dissipated into intrinsic excitations well after scission. The deformation of the fragments at scission is strongly favored by the mutual Coulomb repulsion of the nascent fragments that induces a considerable elongation of their shapes. The energy stored in deformation is transformed into excitation energy of the fragments when they snap back to their ground-state deformation after separation. Therefore, the deformation and collective energies, which largely contribute to the *TXE*, are released well after scission and cannot be exchanged between the fragments. Indeed, there is no possibility to exchange nucleons after scission and practically no mechanism which allows for exchange of energy between the fragments after scission (a small contribution due to Coulomb excitation in the field of the partner fragment may be neglected). Therefore, thermodynamical considerations on the fissioning system as a whole are not meaningful after scission.

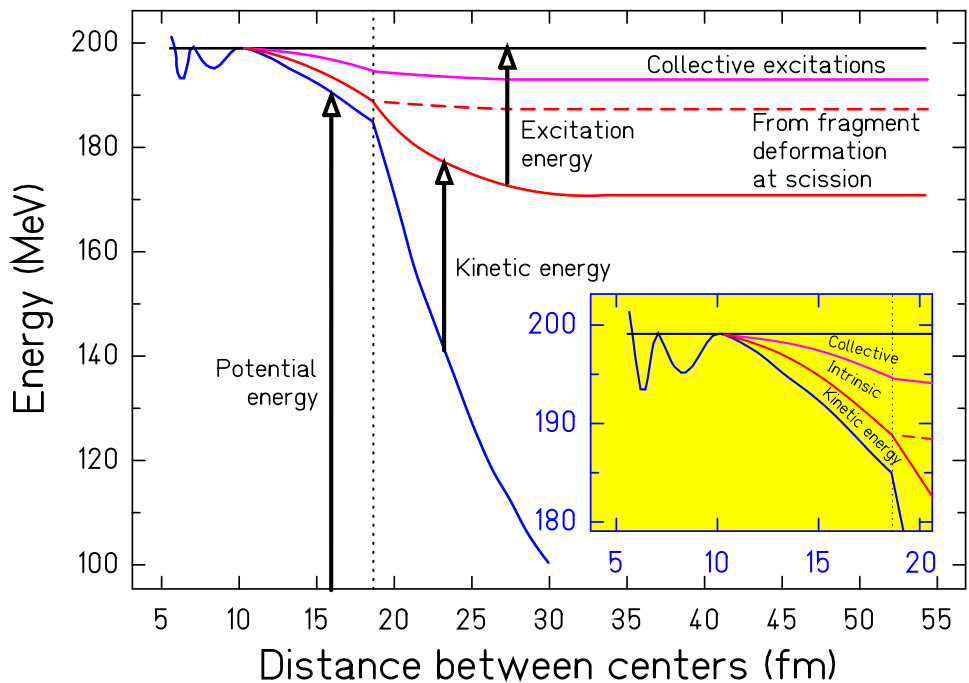

Fig. 1: (Color online) Schematic presentation of the different energies appearing in the fission process. The vertical dotted line indicates the scission point. The inset illustrates that the energy release due to the decreasing potential energy is partly dissipated into excitations of collective normal modes and intrinsic excitations. The remaining part feeds the pre-scission kinetic energy. The main figure demonstrates that the excitation energy of the fragments still increases right after scission, because the excess surface energy of the deformed fragments with respect to their ground states becomes available. Later also the collective excitations are damped into the intrinsic degrees of freedom. The figure represents the fission of $^{236}$U with an excitation energy equal to the fission-barrier height.

**Discussion on existing models**

As has been shown in the previous section, only the intrinsic excitation energy available at scission can be shared between the nascent fragments, before the fragments separate. The total excitation energies of the separated fragments includes other contributions, e.g. from the larger deformations of the fragments in the scission configuration and from the damping of collective excitations after scission. Therefore, it is not appropriate to assume that the total excitation energy of the final fragments is shared between the fragments by the condition of statistical equilibrium. This is at variance with most of the existing models that have been developed for predicting prompt neutron emission from fission fragments [10, 11, 12, 13, 14]. Indeed, these models assume that the *TXE* is distributed among the fragments in a way such that the temperatures of the final fission fragments are equal. Since most often the Fermi Gas level density is used, where excitation energy and temperature are related via $E^*=aT^2$, this leads to:

$$E^*_L/a_L=E^*_H/a_H \qquad (2)$$

where $a_L$ and $a_H$ are the level density parameters of the light (L) and heavy (H) fission fragments. $E^*_L$, $E^*_H$ are the excitation energies of the fully accelerated fission fragments. That is:

$$E^*_L+E^*_H=Q\text{-}TKE=TXE \qquad (3)$$

where Q is the Q-value of the fission reaction and *TKE* is the total kinetic energy. Since neutrons evaporated from the fragments carry only little angular momentum on the average [15] the rotational energy is mostly removed by E2 gamma emission. Therefore, for neutron emission the rotational energy must be subtracted from the *TXE*. This is done in Ref. [14] but it is often forgotten in other models. The temperature values often deduced from Eqs. (2) and (3) by the relation $T=\sqrt{\dfrac{E^*}{a}}$ are a measure of the excitation energies of the final fragments, just transformed into a parameter $T$ with a relation valid for the Fermi gas with some level-density parameter $a$. These temperatures are certainly decisive for the energies of the evaporated neutrons. But neither the excitation energy of the separated fragments nor a somehow deduced temperature has any relevance for the exchange of intrinsic excitation energy between the nascent fragments before scission discussed above. Thus, these temperature values should not be confounded with the temperature values, which govern the energy-sorting mechanism, discussed in [5, 6].

In Refs. [11, 14] it has been shown that a description following eq. (2) does not reproduce experimental data on the average number of neutrons as a function of the fragment mass. To solve this problem a fit parameter $R_T=T_L/T_H$ is introduced [13, 14], in [14] $R_T$ is mass dependent. This parameter serves to somehow "simulate" the contribution of the deformation energy to the final excitation energy of the fission fragments. However, this is again not correct since the *TXE* contains other contributions that are not related to the fragment's deformation ($E^*_{intr}$ and collective excitations). One does not need such a parameter, if the fission process is treated in the way described above: First the intrinsic excitation energy is statistically partitioned between the fragments and then one calculates for each fragment the intrinsic excitation coming from the fragment's deformation and from the damping of collective modes. This is the procedure followed in the GEF code that calculates fission-fragment isotopic yields and neutron spectra from spontaneous fission to an excitation energy of about 13 MeV for a wide range of heavy nuclei from polonium to fermium [16, 17].

**Partition of intrinsic excitation energy: dependence with the level density description**

The understanding of the nuclear level density is still incomplete and under debate. This is mainly because until very recently there has been a great lack of accurate experimental data.

The data were mostly restricted to level counting at low excitation energies and the value at the neutron-separation energy. Therefore, the existing models use a variety of expressions for the level density. Also microscopic models seem to be unable to reproduce recent experimental data, e.g. [4]. More and more precise experiments on level densities of nuclei with masses in the fission-fragment range (e.g. [18]) reveal a "constant temperature" behavior where the logarithm of the level density increases in a good approximation as a linear function of excitation energy, at least in the energy range below the neutron-separation energy. In addition, the experimental work of Ref. [4] shows that the constant-temperature behavior persists up to excitation energies of 20 MeV. In the constant temperature regime, the configuration of the pre-scission di-nuclear system with the largest entropy is characterized by a concentration of all intrinsic excitation energy in the fragment with the smallest logarithmic slope of the energy-dependent level density. Without regard to shell effects, this tends to be the larger of the two fragments. This is what we call the "energy-sorting process" [5], which explains in a transparent way the experimental data of Fig. 1 of Ref. [5] and other similar results that remained a puzzle for decades. It also explains complex data on the dependence of the even-odd effect with mass asymmetry and fissioning nucleus mass [19, 20]. In a recent paper [6] we have discussed possible mechanisms of energy transfer between the fragments like nucleon collisions at the neck or nucleon exchange through the neck.

The combined formula of Gilbert and Cameron [3] is very often used in nuclear-reaction calculations (see pp. 3148-3150 of Ref. [21]) due to its simplicity. Gilbert and Cameron established their composite formula on the basis of experimental indications for a constant-temperature behavior of the nuclear level density well below the neutron separation energy. Above the particle separation threshold, experimental data were practically absent, and the Fermi-gas model was assumed to be valid. This fact led to the creation of the composite Gilbert-Cameron formula consisting of the constant-temperature level density below the particle separation threshold and the Fermi gas level density above the particle separation threshold. The normally used values of this transition energy are in contradiction with Ref. [4] where the constant temperature behavior persists well above the neutron separation energy. It is reasonable to expect a transition between the two regimes. However, it is still not clear up to which excitation energy the constant-temperature component of the Gilbert and Cameron formula is valid.

We performed a numerical calculation to investigate the partition of $E^*_{intr}$ if the nuclear level density is described by the composite Gilbert-Cameron formula. The temperatures as a function of excitation energy are shown in Fig. 2 for the two nuclei $^{94}$Sr and $^{140}$Xe, which are produced with high yields in the thermal-neutron-induced fission of $^{235}$U. We used the global parameterization of the composed Gilbert-Cameron formula recommended in Ref. [21]. Fig. 2 shows that the lowest temperature for which it is possible to have $T_1 = T_2$ is 0.8 MeV. For total excitation energies above 10 MeV, the excitation energies of the two fragments are given by the condition $T_1 = T_2$. For lower total excitation energies there is no solution with $T_1 = T_2$ and the process of thermal equilibration leads to complete energy sorting, see [5]. The result for the excitation energy division at thermal equilibrium is shown in Fig. 3. Several conclusions can be drawn: (i) Complete energy sorting occurs up to about 10 MeV. Thus, energy sorting is not restricted to energies below the matching energies. (ii) Even above the energy range of complete energy sorting, the excitation energy is not divided according to the ratio of the level-density parameters of the two nuclei (see Eq. (2)). The heavier nucleus receives a much larger fraction (we call this "partial energy sorting"). Note that, in the range between 10 and 18 MeV total energy, the equilibrium temperatures of both nuclei stay constant and equal to 0.8 MeV. This is possible because, as can be seen in Fig. 2, from 0 to 8 MeV the light nucleus is in the constant-temperature regime. (iii) Only above 18 MeV, the division of excitation energy can be described without considering the constant-temperature behaviour of the level density below the matching energy. Still, the division of excitation energy deviates from the ratio of the level-density parameters, see [6].

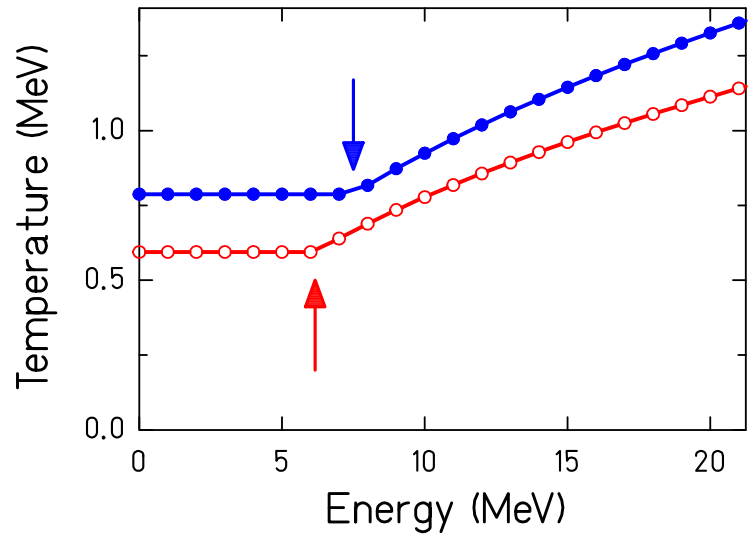

Fig. 2: (Color online) Nuclear-temperature values of $^{94}$Sr (full circles) and $^{140}$Xe (empty circles) as a function of excitation energy according to the composite Gilbert-Cameron formula. The arrows denote the matching energies.

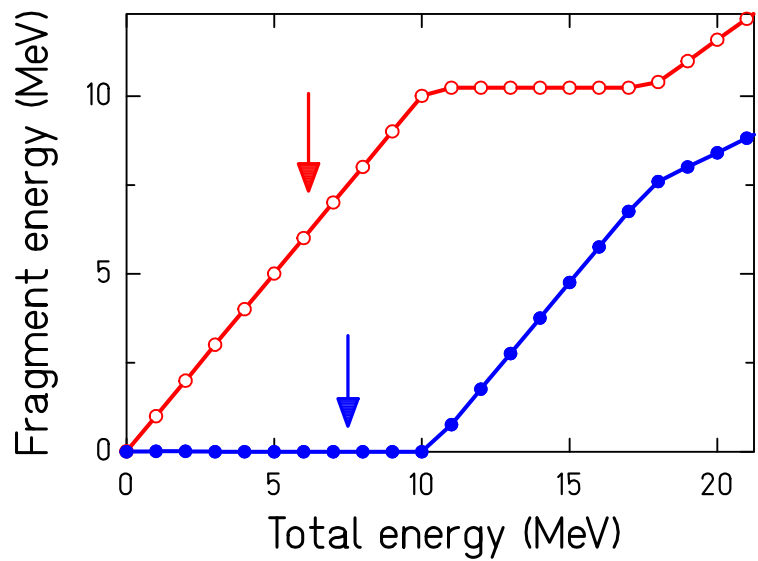

Fig. 3: (Color online) Division of excitation energy between the nuclei $^{94}$Sr (full circles) and $^{140}$Xe (empty circles) in thermal equilibrium as a function of the total excitation energy. The level densities are described by the composite Gilbert-Cameron formula. The arrows denote the matching energies.

Obviously, if the matching energies are shifted to higher values, the region of complete energy sorting will be enlarged and the region of partial energy sorting as well as the region dominated by the Fermi gas description will set in at higher excitation energies.

## Conclusions

We investigated how the final excitation energy in each fission fragment is built up from different contributions that arise at different steps of the fission process. Only the intrinsic excitation energy available at scission is shared between the fragments according to statistical equilibrium: The final partition is the one that maximizes the entropy of the system. The deformation and collective energies are dissipated into intrinsic excitation energy after scission, when the fragments are not in contact anymore. These two types of energies cannot be exchanged between the fragments. Therefore, the assumption of most fission models used for the prediction of prompt neutron emission that the total excitation energy $TXE$ is shared between the fragments according to the condition of equal final temperatures is not correct. We also investigated the intrinsic-excitation-energy partition according to a combined description with a transition from the constant-temperature regime to the Fermi-gas regime. In particular, we made a numerical calculation using the Gilbert-Cameron composite formula. This type of description leads to complete or partial energy sorting at energies well above the matching energy.

### Acknowledgements

This work was supported by the EURATOM 6. Framework Programme "European Facilities for Nuclear Data Measurements" (EFNUDAT), contract number FP6-036434.